\documentclass{ws-ijmpcs}

\begin{document}

\markboth{M.\, Guzzi, P.\, M.\, Nadolsky}
{$Z/\gamma^*$ Transverse momentum distribution at hadron colliders}

%
\catchline{}{}{}{}{}
%

\title{Nonperturbative contributions to 
a resummed leptonic angular distribution in inclusive $Z/\gamma^*$
boson production}

\author{\bf Marco Guzzi\footnote{Presented at the QCD Evolution Workshop,
May 14-17 2012, Thomas Jefferson National Accelerator Facility, Newport News, VA}, 
Pavel M. Nadolsky}



\address{Department of Physics, Southern Methodist University, 3215 Daniel Avenue \\
Dallas, Texas 75275,
USA\\
mguzzi@physics.smu.edu, nadolsky@physics.smu.edu}

\maketitle

\begin{history}
\received{15 August 2012}
\revised{Day Month Year}
\end{history}

\begin{abstract}
We summarize a new analysis of the distribution $\phi_{\eta}^{*}$
of charged leptons produced in decays of $Z$ and $\gamma^{*}$ bosons
in the Collins-Soper-Sterman (CSS) formalism for transverse momentum
resummation. By comparing the $\phi_{\eta}^{*}$ distribution measured
at the Tevatron with the resummed CSS cross section with approximate
${\cal O}(\alpha_{s}^{2})$ Wilson coefficients, we constrain the
magnitude of the nonperturbative Gaussian smearing factor and analyze
its uncertainty caused by variations in scale parameters. We find
excellent agreement between the $\phi_{\eta}^{*}$ data and our theoretical
prediction, provided by the \textsc{ResBos} resummation program. The nonperturbative
factor that we obtained can be used to update resummed QCD predictions
for precision measurements in inclusive $W$ and $Z$ production and
for comparisons to various models of nonperturbative dynamics.

\keywords{Transverse momentum resummation; QCD Factorization}
\end{abstract}

\ccode{PACS numbers: 11.25.Hf, 123.1K}

\vspace{0.8cm}

{\bf Leptonic angular distribution $\phi^*_\eta$ as a probe of nonperturbative dynamics.} 
Precision of the current data from the Tevatron and LHC on the inclusive transverse momentum ($Q_T$) 
distributions of $W$ and $Z$ bosons imposes
growing demands on theoretical calculations. $Q_T$ distributions 
are used for lucent tests of QCD factorization 
and in measurements of the $W$ boson mass that place important constraints 
on the parameters of electroweak symmetry breaking. A variety of 
radiative contributions affect the $Q_T$ distribution of a heavy boson 
at the current level of accuracy. They include NNLO QCD and NLO EW perturbative 
corrections, logarithmic QCD contibutions that dominate $d\sigma/dQ_T$ 
when $Q_T\rightarrow 0$, and also nonperturbative power-suppressed terms 
that modify the $Q_T$ distribution when $Q_T$ is below a few GeV.
 Collins, Soper, and Sterman (CSS) [\refcite{Collins:1981uk,Collins:1981va,Collins:1984kg}]
have developed a QCD factorization approach to include all such terms order by order in $\alpha_s$.
The CSS formalism combines the resummation of large Sudakov factors in the small-$Q_T$ limit 
[\refcite{Dokshitzer:1978yd,Parisi:1979se,Curci:1979bg}] 
with fixed-order QCD contributions at large $Q_T$. As a recent development, 
the NNLL/NNLO expression for the resummed $Q_T$ distribution has been published [\refcite{Bozzi:2010xn}]. 
The CSS method is a classical realization of the factorization based on 
transverse-momentum dependent (TMD) distribution functions, in which 
PDFs and fragmentation functions depend explicitly on intrinsic transverse momentum 
in addition to the usual momentum fraction variables. While theoretical methods of 
TMD factorization  [\refcite{Aybat:2011zv,GarciaEchevarria:2011rb,Echevarria:2012pw,Collins:1999dz,Collins:2000gd,Henneman:2001ev,Belitsky:2002sm,Boer:2003cm,Collins:2003fm,Collins:2007ph,Cherednikov:2007tw,Cherednikov:2009wk}] 
and soft-collinear effective theory [\refcite{GarciaEchevarria:2011rb,Becher:2011xn,Aybat:2011zv,Mantry:2010mk,Mantry:2010bi,Idilbi:2005er}]
undergo rapid developments, the CSS formalism is well-suited for detailed phenomenological studies, as it is implemented 
in detail in practical simulations. In this short paper, we summarize a recent application of the
CSS formalism for the computation of the angular distribution $d\sigma/d\phi^*_\eta$ of
the lepton pairs in  $Z/\gamma^*$ boson production at the Tevatron. Full details 
will be presented in a separate forthcoming paper [\refcite{GNW}]. 
The $\phi^*_\eta$ distribution is closely related to the $Q_T$ distribution 
in the limit $\phi^*_\eta \propto Q_T/Q\rightarrow 0$. Its analysis may provide 
valuable insights about the nonperturbative QCD dynamics.

The D\O\, collaboration has published detailed measurements of the $\phi^*_\eta$ dependence 
in the electron and muon decay channels [\refcite{Abazov:2010mk}]. 
We ask if the D\O\, data corroborate the universal behavior of the resummed nonperturbative terms 
that was observed in the global analyses of $Q_T$ distributions 
in  $\gamma^*$ and $Z$ production at fixed-target and collider energies [\refcite{Landry:2002ix,Konychev:2005iy}].
We also investigate the rapidity dependence of the nonperturbative terms, which may be indicative of 
new types of higher-order contributions [\refcite{Berge:2004nt}].
The small-$Q_T$ part of the distribution is obtained in the CSS formalism 
by the Fourier-Bessel transform of a form factor $\widetilde W(b,Q)$ that depends on 
the transverse position variable $b$. For large boson virtualities $Q$ of order 100 GeV, 
the overall form of the $Q_T$ distribution {\it at all $Q_T$} is determined by small-$b$ contributions, 
arising at energy scales $\mu \sim 1/b \gg 1$ GeV. 
Such contributions are entirely predicted in perturbative QCD theory. They dominate the cross section
at any $Q_T$ value [\refcite{Parisi:1979se,Arnold:1990yk}]. 
When $Q_T$ is below 5 GeV, the production rate is also mildly sensitive to the behavior of $\widetilde W(b,Q)$ 
at $b> 0.5$ GeV$^{-1}$, where the perturbative expansion is increasingly unreliable because of the 
vicinity of the Landau pole in $\alpha_s(1/b)$. We are interested to know which forms of the large-$b$ 
extrapolation of  $\widetilde W(b,Q)$  are compatible with the observed behavior 
of $Q_T$ and  $\phi^*_\eta$ distributions.
A comprehensive solution for $\widetilde W(b,Q)$ at large $b$ remains elusive, 
but efforts to derive it produced several instructive models, such as [\refcite{Korchemsky:1994is,Ellis:1997ii,Guffanti:2000ep,Qiu:2000hf,Tafat:2001in,Kulesza:2002rh,Becher:2011xn}]. 
From the phenomenological point of view, 
a typical $Z$ data set does not have sensitivity to distinguish between these models. 
In deeply nonperturbative region that is characterized by $b\gtrsim$ 1 GeV$^{-1}$, the form factor 
$\widetilde W(b,Q)$ is strongly suppressed and does not contribute to $d\sigma/dQ_T$ for $Q$ of order $M_Z$
[\refcite{Konychev:2005iy}]. Only the contributions from the transition region 
of $b$ of about $1$ GeV$^{-1}$, where $\mu\approx 1/b$ is about 1 GeV, 
are numerically non-negligible compared to the leading-power cross section predicted by perturbative QCD theory.
In the transition region, the extrapolation of the perturbative expression provides a reasonable approximation 
for the leading-power (logarithmic) part of $\widetilde W(b,Q)$. It can be realized in the 
``revised $b_*$ model'' [\refcite{Konychev:2005iy}], which parametrizes the extrapolated contribution by a flexible form that depends on a single parameter $b_{max}$. In addition, a few suppressed terms proportional 
to even powers of $b$ play some role in this interval of $b$. In comparisons to the experimental data, 
their cumulative effect can be usually approximated  by a single Gaussian smearing factor  $\exp\{-a(Q) b^2\}$, 
where the coefficient $a(Q)$ is found from the experiment. 
This arrangement provides a few-parameter approximation for viable nonperturbative models 
in the phenomenologically relevant region of $b$. 
In our previous work [\refcite{Konychev:2005iy}], the magnitude and $Q$-dependence of the Gaussian factor 
were determined from $Q_T$ distributions of the Drell-Yan pairs. Recently, 
the $\phi^*_\eta$ distribution has been proposed as a sensitive probe of the small-$Q_T$ dynamics 
[\refcite{Banfi:2010cf}], as it has reduced uncertainties associated with the lepton momentum resolution. 
Here we update the constraints on the Gaussian smearing factor in $Z$ boson production using the $\phi^{*}_{\eta}$ distribution.

Our analysis is carried out using the program \textsc{ResBos} [\refcite{Balazs:1997xd,Landry:2002ix,RESBOSweb}], 
which realizes the CSS formalism to compute  fully differential cross sections of lepton pairs 
in production of high-mass virtual photons ($\gamma^*$) and heavy electroweak bosons ($W$, $Z$, and $H$). 
A resummed treatment of new variables $a_T$ and $\phi^*_\eta$ and their relationship 
to $Q_T$ was studied in [\refcite{Banfi:2012du,Banfi:2011dx}]. The fully differential output from \textsc{ResBos} 
can also be cast in the form of the $\phi^*_\eta$ distribution. In the new analysis, we 
find an excellent agreement between the \textsc{ResBos} prediction and the $\phi^*_\eta$ data, contrary to the conclusion 
made by the D\O\, paper  [\refcite{Abazov:2010mk}]. However, the quality of agreement depends on the inclusion 
of perturbative loop contributions and selection of scales in the resummed cross section. 
In Refs.~[\refcite{Banfi:2012du,Banfi:2011dx,Banfi:2011dm,Banfi:2009dy,Bozzi:2010xn}], 
the evidence for a nonzero nonperturbative factor in $Z$ production 
has been contested in the light of the uncertainty in the resummed cross section 
due to the dependence on factorization scales.  
The evidence for nonperturbative smearing is inconclusive at the NLL+NLO level because of a large scale dependence.
To address this point, we fully include the scale dependence in the small-$Q_T$ part 
of the resummed cross section up to ${\cal O}(\alpha_s^2)$ [\refcite{GNW}]. Despite the scale uncertainties, 
the statistical analysis of the fit to the $\phi^*_\eta$ 
data indicates pronounced preference for a nonperturbative Gaussian contribution with $a(M_Z)\approx 1$ GeV$^{2}$.
The main reason is that the Gaussian suppression of the large-$b$ tail of $\widetilde W(b,Q)$ alters the resummed $d\sigma/dQ_T$ in a different way than the factorization scales in the leading-power part of $\widetilde W(b,Q)$. The nonperturbative Gaussian factor suppresses the rate only at $Q_T$ below 2-3 GeV, 
while the leading-power scale dependence affects a broader interval of $Q_T$ values. This results in a characteristic shift of the peak in the $d\sigma/dQ_T$ distribution due to the nonperturbative suppression, which is distinct from the typical scale dependence. 

\begin{figure}[ht]
 \centerline{\psfig{file=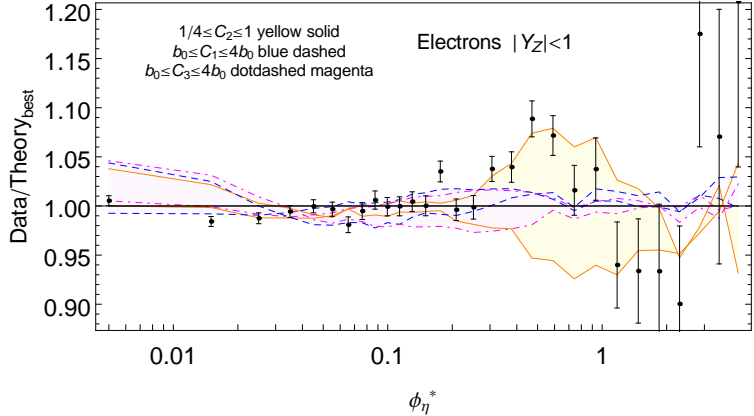,width=10cm}}
\vspace*{8pt}
 \caption{Impact of the scale variation on the agreement between theory and
data for $|y_{Z}|\leq1$. \label{f1}}
\end{figure}

{\bf Particulars of the calculation.} 
In the CSS formalism,  the full resummed (RES) $Q_T$ distribution is commonly 
represented as a combination of the resummed (W), finite-order (FO), and asymptotic (ASY) terms [\refcite{Collins:1984kg}]:
\begin{equation}
\left(\frac{d\sigma}{dQ_T^2}\right)_{RES} = \left(\frac{d\sigma}{dQ_T^2}\right)_{W} + \left(\frac{d\sigma}{dQ_T^2}\right)_{FO}
- \left(\frac{d\sigma}{dQ_T^2}\right)_{ASY}.
\end{equation} 
The accuracy of the D\O\, measurement demands that the main perturbative 
corrections due to the QCD and electroweak radiation are included. Numerical, but
not analytical, results for the complete NNLO QCD contribution to the resummed $Q_T$ distribution in $Z/\gamma^*$ production at the Tevatron have been recently released [\refcite{Bozzi:2010xn}]. \textsc{ResBos} includes the dominant NNLO QCD contributions and provides a faithful estimate for the remaining small NNLO contribution that has not been published in an analytical form, as summarized below. Thus, effectively \textsc{ResBos} is close to the full NNLO precision in the kinematical region relevant for this analysis.
The electroweak (EW) corrections to  $Z$ production compete in magnitude with NNLO QCD contributions and are
available to NLO [\refcite{Baur:2001ze,Zykunov:2005tc,CarloniCalame:2007cd,Arbuzov:2007db}]. 
In the comparison to the D\O\, data, we do not include the EW corrections in our theory prediction,
but correct the fitted $\phi^*_\eta$ data by subtracting the predominant correction 
due to the final-state photon radiation obtained by the \textsc{Photos} code [\refcite{Barberio:1993qi}]. 
Upon the inclusion of these contributions, scale dependence remains a major 
systematic uncertainty affecting our theory prediction. 
In the latest \textsc{ResBos} implementation, the fully differential CSS resummed
cross section is given by {[}\refcite{Balazs:1997xd}{]} \begin{eqnarray}
 &  & \frac{d\sigma\left(h_{1}h_{2}\rightarrow(Z\rightarrow\ell\bar{\ell})X\right)}{dQ^{2}dydQ_{T}^{2}d\Omega}=\frac{1}{48\pi S}\frac{Q^{2}}{(Q^{2}-M_{Z}^{2})^{2}+Q^{4}\Gamma_{Z}^{2}/M_{Z}^{2}}\nonumber \\
 &  & \times\Biggl\{\int\frac{d^{2}b}{4\pi^{2}}\, e^{i\vec{Q}_{T}\cdot\vec{b}}\sum_{j=u,d,s...}\tilde{W}_{j}^{pert}(b_{*},Q,x_{1},x_{2},\Omega,C_{1},C_{2},C_{3})\tilde{W}^{NP}(b,Q)\nonumber \\
 &  & +Y(Q_{T},Q,x_{1},x_{2},\Omega,C_{4})\Biggr\},\label{resummed1}\end{eqnarray}
 in notations of Ref.~{[}\refcite{Balazs:1997xd}{]}. The leading-power
({}``perturbative'') form factor $\tilde{W}^{pert}$ is defined
by \begin{eqnarray}
\tilde{W}^{pert} & = & \sum_{j=u,d,s...}\left|H(Q,\Omega,C_{4})\right|^{2}\,\exp\left[-\int_{C_{1}^{2}/b^{2}}^{C_{2}^{2}Q^{2}}\frac{d\bar{\mu}^{2}}{\bar{\mu}^{2}}A(\bar{\mu};C_{1})\,\ln\left(\frac{C_{2}^{2}Q^{2}}{\bar{\mu}^{2}}\right)+B(\bar{\mu};C_{1},C_{2})\right]\nonumber \\
 & \times & \sum_{a=g,q}\left[{\cal C}_{ja}\otimes f_{a/h_{1}}\right]\left(x_{1},\frac{C_{1}}{C_{2}},\frac{C_{3}}{b}\right)\sum_{b=g,q}\left[{\cal C}_{\bar{j}b}\otimes f_{b/h_{2}}\right]\left(x_{2},\frac{C_{1}}{C_{2}},\frac{C_{3}}{b}\right),\label{defW}\end{eqnarray}
in terms of the hard part $H(Q,\Omega,C_{4})$ (dependent on $Q$
and the solid angle $\Omega=\left\{ \theta_{*},\varphi_{*}\right\} $
of $Z$ boson decay in the Collins-Soper frame), Sudakov exponent,
and convolutions $\left[{\cal C}_{j/a}\otimes f_{a/H}\right]$ of
Wilson coefficient functions ${\cal C}_{j/a}(z,\, C_{1}/C_{2},\,\mu_{F}=C_{3}/b)$
and parton distribution function $f_{a/H}(z,\mu_{F})$ for a parton
$a$ inside the initial-state hadron $h$. $\tilde{W}^{NP}(b,Q)$
is the nonperturbative factor explained below. $Y$ is the difference
between the finite-order and asymptotic cross sections and it dominates at
$Q_{T}$ of order $Q$. 
Constants $C_{i}$ (for $i=1,...,4)$ are the scale coefficients that
determine several factorization scales introduced by resummation.
They arise because $\tilde{W}^{pert}$ depends on two distinct momentum
scales $1/b$ and $Q$. Specifically, $C_{1}=b \mu$ and $C_{2}=\mu/Q$,
where $\mu$ is the scale introduced by the evolution equations that
control the large logarithms. The constant $C_{3}$ arises in the
Wilson covolutions$\left[{\cal C}_{j/a}\otimes f_{a/H}\right]$ and
specifies the factorization scale $\mu_{F}=C_{3}/b$ in the PDFs $f_{a/H}(z,\mu_{F})$.
$C_{4}=\mu_{H}/Q$ is the scale constant in the hard part $H$ of
the resummed term $W$ and the $Y$ piece, which we select  as $C_{4}=C_{2}$
for simplicity. The expression for the resummed cross section becomes
particularly simple for a {}``canonical'' combination of the scale
coefficients, given by $C_{1}=b_{0},\, C_{2}=1,\, C_{3}=b_{0}$, where
$b_{0}=2e^{-\gamma_{E}}$, and $\gamma_{E}=0.577...$ is the Euler\textendash{}Mascheroni
constant. 
We evaluate $Y$ to ${{\cal O}}(\alpha_{s}^{2})$ based on the calculation
in  [\refcite{Arnold:1989ub,Arnold:1988dp,Melnikov:2006kv}]. The
functions $A(\bar{\mu};C_{1})$ and $B(\bar{\mu};C_{1},C_{2})$ are
computed up to orders ${\cal O}(\alpha_{s}^{3})$ and ${O}$($\alpha_{s}^{2})$,
respectively. The Wilson coefficient functions are computed exactly
up to ${{\cal O}}(\alpha_{s}).$ The only unavailable part of the
NNLO resummed cross section is the ${\cal O}(\alpha_{s}^{2})$ Wilson
coefficient ${{\cal C}}_{ja}^{(2)}$ which receives contributions from two loop virtual diagrams.
However, from the fixed order NNLO calculation [\refcite{Melnikov:2006kv}] this contribution 
is small in magnitude (2-3\%), mostly affects the overall normalization of the $W$ term,
and has weak kinematical dependence.
Thus, without losing accuracy, one can approximate this term by \begin{equation}
{{\cal C}}_{ja}^{(2)}(z,C_{1}/C_{2},C_{3})=\delta {\cal C}^{(2)}\,\delta(1-z)\,\delta_{ja}+L(C_{1},C_{2},C_{3}),\end{equation}
where $L(C_{1},C_{2},C_{3})=0$ for the canonical combination. Since
$\delta {\cal C}^{(2)}$ is nearly constant in the kinematical region relevant
for $Z$ production, we estimate its magnitude from the known value
of the $O(\alpha_{s}^{2})$ $K$-factor for the inclusive cross section
$d\sigma/dQ$ that is known for a long time {[}\refcite{Hamberg:1990np}{]}
and was evaluated in our analysis by the computer code \textsc{Candia}
{[}\refcite{Cafarella:2007tj,Cafarella:2008du}{]}. The inclusion
of the estimated $\delta C^{(2)}$ in the calculation has practically
no effect on our conclusions. Finally, $L(C_{1},C_{2},C_{3})$ is
found \emph{exactly} by requiring the independence of the $\alpha_{s}$
series expansion of $\tilde{W}$ on the choice of $C_{1},$ $C_{2},$
and $C_{3}$ order by order. By truncating the series at ${\cal O}(\alpha_{s}^{2})$,
we must have the equality 
$\tilde{W}(b,Q,C_{1},C_{2},C_{3})\vert_{{\cal O}(\alpha_{s}^{2})}=\tilde{W}(b,Q,C_{1}=C_{3}=b_{0},C_{2}=1)\vert_{{\cal O}(\alpha_{s}^{2})}$,
which allows us to completely reconstruct the dependence of the ${\cal O}(\alpha_{s}^{2})$
part on the scale parameters $C_{i}$.  
The dependence on the scale constants is illustrated in Fig.~\ref{f1},
which shows the ratio of the experimental and best-fit theoretical
values of $(1/\sigma)\cdot d\sigma/d\phi_{\eta}^{*}$ for electrons
with a constraint $\left|y_{Z}\right|\leq 1$ on the $Z$ boson rapidity.
The best agreement with the data are obtained for $\{C_{1}=2b_{0},\, C_{2}=1/2,\, C_{3}=2b_{0}\}$.
Several error bands are obtained by variations of the indicated scale
parameters around this best-fit combination. The variation of the
scales affects the quality of the fit quantified by $\chi^{2}$ and
modifies the cross section in a wide range of $\phi_{\eta}^{*}$.
In contrast, the variation of the nonperturbative factor is 
pronounced only at $\phi_{\eta}^{*}\lesssim0.5$, which corresponds to typical $Q_{T}$
of a few GeV. This difference allows one to discriminate between the
nonperturbative $Q_{T}$ smearing and perturbative scale dependence.

{\bf The nonperturbative factor.}
In order to extrapolate $\widetilde{W}^{pert}$ in Eq.~(\ref{resummed1})
to the large $b$ values of order or above $1\mbox{ GeV}^{-1}$, we evaluate
it in the revised $b_{*}(b,b_{max})$ model {[}\refcite{Konychev:2005iy}{]}
as a function of $b_{*}\equiv b/\sqrt{1+(b/b_{max})^{2}}$ with $b_{max}=1.5\mbox{ GeV}^{-1}$.
The factorization scale in the convolutions $[C\otimes f]$ is set in this model to $\mu_F=C_3/b_{*}(b,b_0/Q_0)$, where
$Q_0=1$ GeV is the initial scale of the PDFs. This choice of $b_{max}$ is preferred by the global fit to Drell-Yan
$Q_{T}$ data, where it both improves the agreement with the data
and preserves the exact form of the perturbative expansion for $\widetilde{W}^{pert}$
at $b<1\mbox{ GeV}^{-1}$.
The $\phi_{\eta}^{*}$ data is dominated by the narrow vicinity
of $Q$ around $M_{Z}.$ Therefore, for the purpose of this analysis,
it sufficies to approximate the nonperturbative Gaussian factor as
\begin{equation}
\tilde{W}^{NP}\left(b,Q\approx M_{Z}\right)=\textrm{exp}\left[-a_{1}(M_{Z})\, b^{2}\right]\,.\label{WNPGNW}\end{equation}
This expression simplifies a more general parametrization {[}\refcite{Konychev:2005iy,Landry:2002ix}{]}
\begin{eqnarray}
\tilde{W}^{NP}\left(b,Q\right)=\textrm{exp}\left[-b^{2}\left(a_{1}+a_{2}\ln\left(\frac{Q}{Q_{0}}\right)+a_{3}\ln\left(\frac{x_{1}x_{2}}{0.01}\right)\right)\right]\,,\label{KN}\end{eqnarray}
in which the coefficients $a_{1},$ $a_{2,}$and $a_{3}$ can be separated
by fitting to several data sets at distinct $\sqrt{s}$ and $Q$ combinations.
Once $a_{1}(M_Z)$ is known, the coefficients $a_{2}$ and
$a{}_{3}$ from the global $Q_{T}$ fit of Ref.~{[}\refcite{Konychev:2005iy}{]}
can be used to find $\tilde{W}^{NP}$ in $W$ boson production.
\begin{figure}[ht]
 \centerline{\psfig{file=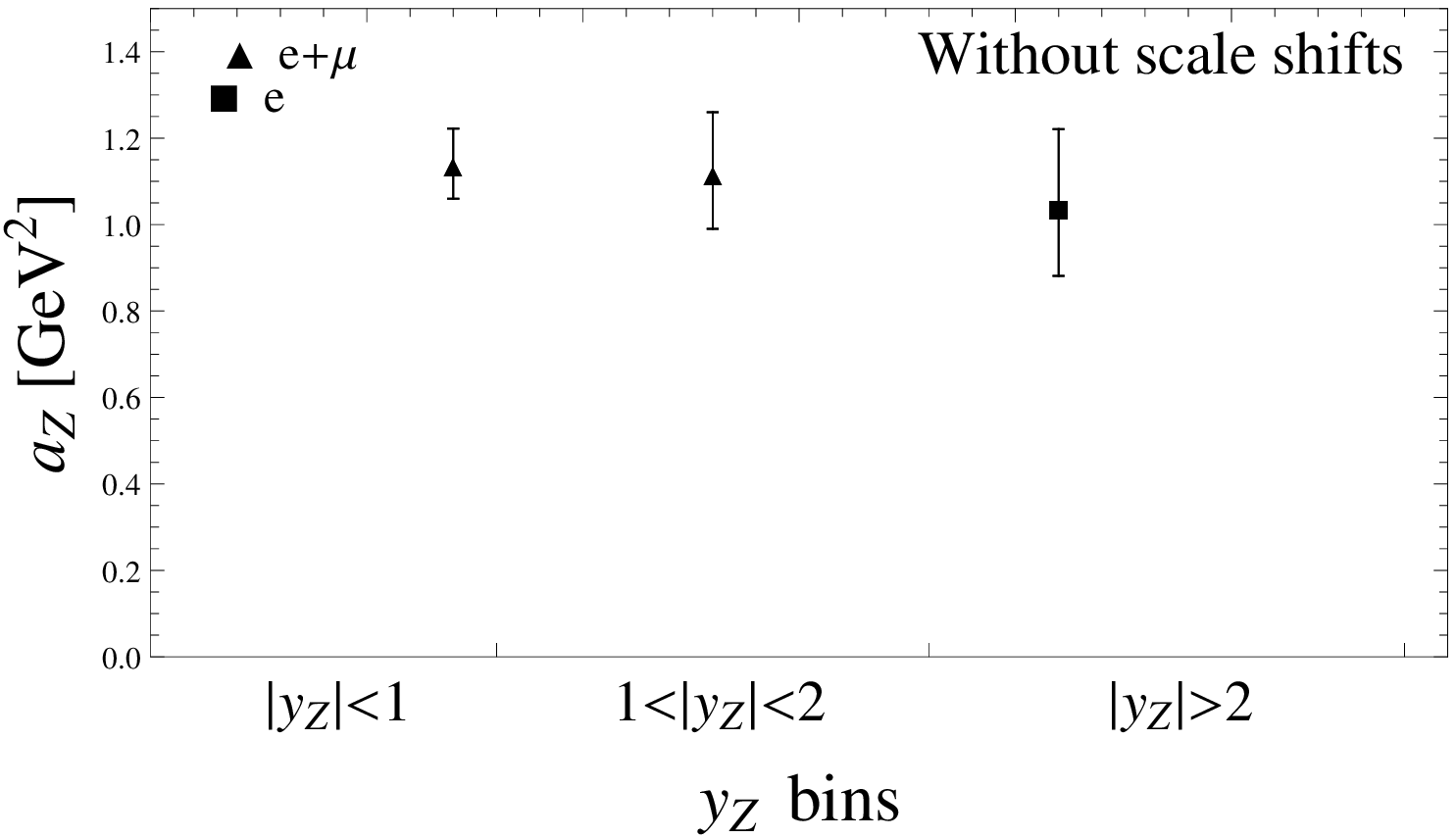,width=2.9in,height=120pt} \psfig{file=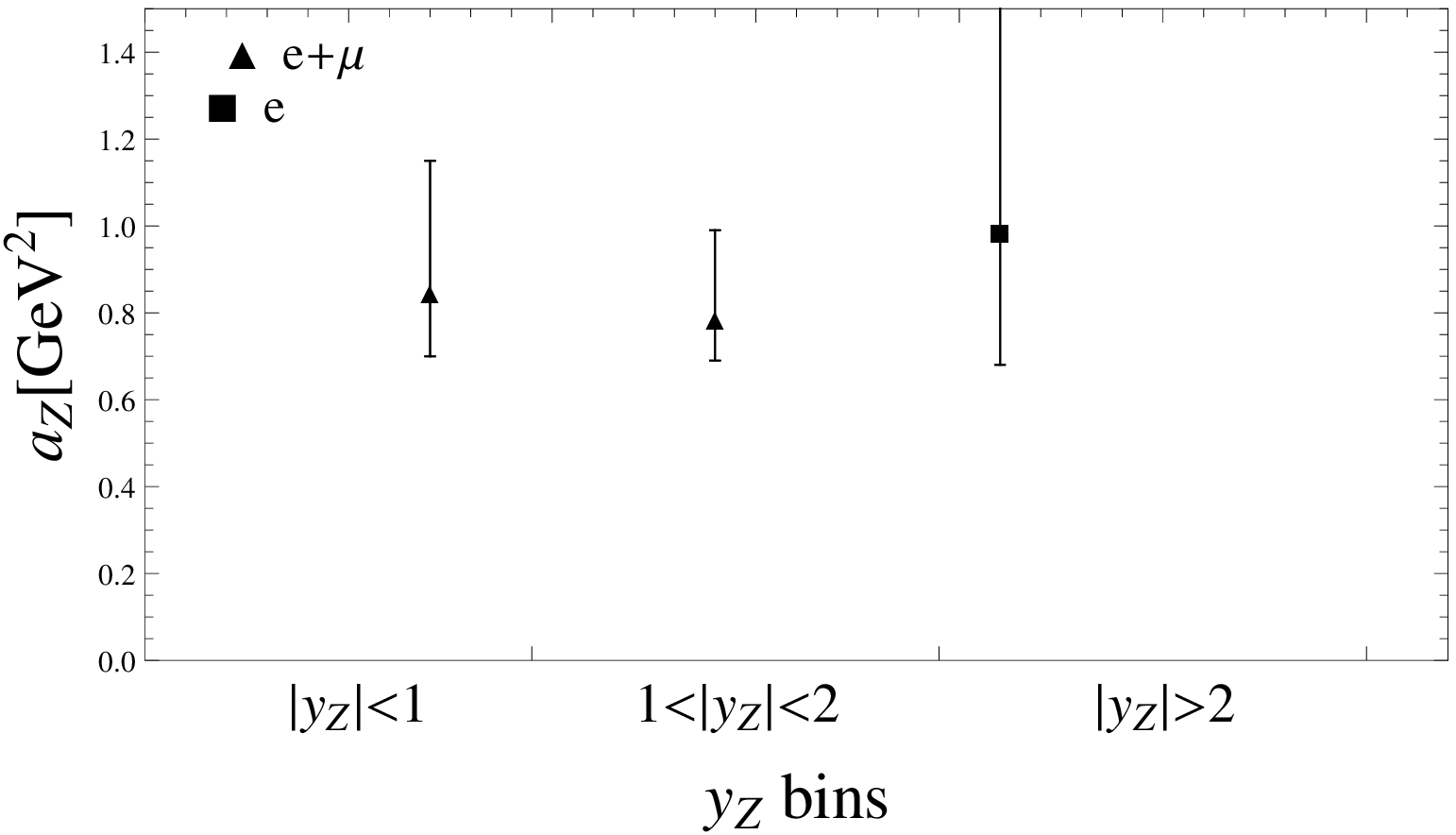,width=2.9in,height=120pt}}
\vspace*{8pt}
 \caption{Estimate of coefficient $a_{1}$ in the three bins of rapidity of
the data without (left inset) and with (right inset) $C_{1},C_{2}$
and $C_{3}$ shifts.\label{f2}}
\end{figure}
In each $y_Z$ bin of the electron and muon $\phi_\eta^*$ data, we compute 
$\chi^2$ and use it to determine the nonperturbative coefficient $a_1$. The
results are shown in Fig.~\ref{f2}. In order to estimate the impact of 
the scale dependence, the fit includes a correlation matrix quantifying 
the uncertainty in the theory cross section due to variations in the 
scale constants $C_{1},C_{2}$ and $C_{3}$. In this case, we assign the 
68\% confidence to the interval $-1\leq \log_2(C_i/C_i^{best-fit})\leq 1$.
According to the figure, all rapidity bins generally prefer a non-zero $a_1$, even
if the scale shifts are included. The scale dependence increases the errors and makes 
them very asymmetric, but a downward change in the best-fit $a_1$ is more disfavored than the upward change. 
With the scale shifts, we obtain the central value of  $a_{1}(M_Z)=0.82\pm{0.12}$ GeV$^{2}$. 
No significant dependence of the best-fit $a_1(M_Z)$ values on the rapidity $y_Z$ is observed,
but the uncertainty is very large in the $|y_Z|>2$ bin.

{\bf Conclusions.}
A new version of the resummation program \textsc{ResBos}
was employed to examine the differential cross section at small values of the leptonic angle $\phi^*_\eta$
at the Tevatron. The data on the $Z/\gamma^*$ $\phi^*_\eta$ distribution
collected by the D\O\, collaboraton was compared with the CSS resummed cross sections
with approximate ${\cal O}(\alpha_s^2)$ Wilson coefficient functions and complete  ${\cal O}(\alpha_s^2)$
scale dependence. \textsc{ResBos} agrees well with these data, provided we use the hard scales $Q/2$ (i.e.,
the scale coefficients $C_2=C_4=1/2$) in the resummed cross section. We determined 
the nonperturbative factor $\widetilde{W}^{NP}(b,Q)$ preferred by the $\phi^*_\eta$, while evaluating 
the effects that have comparable magnitude: QCD scale dependence at ${\cal O}(\alpha_s^2)$ 
and NLO electromagnetic contributions. In a fit that allowed for variations of the resummation 
scale parameters $C_i$, we observe a $2.5\sigma$ preference 
for a non-zero Gaussian smearing factor $a_1(M_Z)\approx 1$ GeV$^{2}$. This central value is in agreement with the findings
of the previous analyses of $Z$ $Q_T$-dependent distributions [\refcite{Qiu:2000hf,Kulesza:2002rh,Konychev:2005iy}]. 
The non-pertubative factor $\widetilde{W}^{NP}(b,Q)$ is a part of the complete resummed factor 
$\widetilde{W}^{pert}(b_*,Q)\widetilde{W}^{NP}(b,Q)$, which is well-constrained in the phenomenologically relevant
region $b\lesssim 1\mbox{ GeV}^{-1}$ by a combination of the ${\cal O}(\alpha_s^2)$ PQCD calculation 
and the nonperturbative factor found from the experimental data. 
Further information on this study will be provided in Ref.~[\refcite{GNW}]. It will be used to
update \textsc{ResBos} predictions for future Tevatron and LHC studies.

\vspace{-0.2cm}
\section*{Acknowledgments}
M.G. would like to thank the organizers of ``The QCD Evolution workshop 2012'' and C.-P. Yuan for related discussions.
This work was supported by the U.S. DOE Early Career Research Award DE-SC0003870
and by the Lightner Sams Foundation.

\bibliographystyle{h-elsevier3}

\begin{thebibliography}{10}

\bibitem{Collins:1981uk}
J.C. Collins and D.E. Soper,
\newblock Nucl.Phys. B193 (1981) 381.

\bibitem{Collins:1981va}
J.C. Collins and D.E. Soper,
\newblock Nucl.Phys. B197 (1982) 446.

\bibitem{Collins:1984kg}
J.C. Collins, D.E. Soper and G.F. Sterman,
\newblock Nucl.Phys. B250 (1985) 199.

\bibitem{Dokshitzer:1978yd}
Y.L. Dokshitzer, D. Diakonov and S. Troian,
\newblock Phys.Lett. B79 (1978) 269.

\bibitem{Parisi:1979se}
G. Parisi and R. Petronzio,
\newblock Nucl.Phys. B154 (1979) 427.

\bibitem{Curci:1979bg}
G. Curci, M. Greco and Y. Srivastava,
\newblock Nucl.Phys. B159 (1979) 451.

\bibitem{Bozzi:2010xn}
G. Bozzi et~al.,
\newblock Phys.Lett. B696 (2011) 207.

\bibitem{Aybat:2011zv}
S.M. Aybat and T.C. Rogers,
\newblock Phys.Rev. D83 (2011) 114042.

\bibitem{GarciaEchevarria:2011rb}
M. Garcia-Echevarria, A. Idilbi and I. Scimemi,
\newblock JHEP 1207 (2012) 002.

\bibitem{Echevarria:2012pw}
M. Garcia-Echevarria, A. Idilbi, A. Schafer and I. Scimemi,
\newblock arXiv:1208.1281[hep-ph].


\bibitem{Collins:1999dz}
J.C. Collins and F. Hautmann,
\newblock Phys.Lett. B472 (2000) 129.

\bibitem{Collins:2000gd}
J.C. Collins and F. Hautmann,
\newblock JHEP 0103 (2001) 016.

\bibitem{Henneman:2001ev}
A. Henneman, D. Boer and P. Mulders,
\newblock Nucl.Phys. B620 (2002) 331.

\bibitem{Belitsky:2002sm}
A.V. Belitsky, X. Ji and F. Yuan,
\newblock Nucl.Phys. B656 (2003) 165.

\bibitem{Boer:2003cm}
D. Boer, P. Mulders and F. Pijlman,
\newblock Nucl.Phys. B667 (2003) 201.

\bibitem{Collins:2003fm}
J.C. Collins,
\newblock Acta Phys.Polon. B34 (2003) 3103.

\bibitem{Collins:2007ph}
J. Collins, T. Rogers and A. Stasto,
\newblock Phys.Rev. D77 (2008) 085009.

\bibitem{Cherednikov:2007tw}
I. Cherednikov and N. Stefanis,
\newblock Phys.Rev. D77 (2008) 094001.

\bibitem{Cherednikov:2009wk}
I. Cherednikov and N. Stefanis,
\newblock Phys.Rev. D80 (2009) 054008.

\bibitem{Becher:2011xn}
T. Becher, M. Neubert and D. Wilhelm,
\newblock JHEP 1202 (2012) 124.

\bibitem{Mantry:2010mk}
S. Mantry and F. Petriello,
\newblock Phys.Rev. D83 (2011) 053007.

\bibitem{Mantry:2010bi}
S. Mantry and F. Petriello,
\newblock Phys.Rev. D84 (2011) 014030.

\bibitem{Idilbi:2005er}
A. Idilbi, X. Ji and F. Yuan,
\newblock Phys.Lett. B625 (2005) 253.

\bibitem{GNW}
M. Guzzi, P.M. Nadolsky and B. Wang,
\newblock (in preparation).

\bibitem{Abazov:2010mk}
D\O\, Collaboration, V.M. Abazov et~al.,
\newblock Phys.Rev.Lett. 106 (2011) 122001.

\bibitem{Landry:2002ix}
F. Landry, R. Brock, P.M. Nadolsky and C.-P. Yuan,
\newblock Phys.Rev. D67 (2003) 073016.

\bibitem{Konychev:2005iy}
A.V. Konychev and P.M. Nadolsky,
\newblock Phys.Lett. B633 (2006) 710.

\bibitem{Berge:2004nt}
S. Berge, P.M. Nadolsky, F. Olness and C.-P. Yuan,
\newblock Phys.Rev. D72 (2005) 033015.

\bibitem{Arnold:1990yk}
P.B. Arnold and R.P. Kauffman,
\newblock Nucl.Phys. B349 (1991) 381.

\bibitem{Korchemsky:1994is}
G.P. Korchemsky and G.F. Sterman,
\newblock Nucl.Phys. B437 (1995) 415.

\bibitem{Ellis:1997ii}
R.K. Ellis and S. Veseli,
\newblock Nucl.Phys. B511 (1998) 649.

\bibitem{Guffanti:2000ep}
A. Guffanti and G. Smye,
\newblock JHEP 0010 (2000) 025.

\bibitem{Qiu:2000hf}
J. Qiu and X. Zhang,
\newblock Phys.Rev. D63 (2001) 114011.

\bibitem{Tafat:2001in}
S. Tafat,
\newblock JHEP 0105 (2001) 004.

\bibitem{Kulesza:2002rh}
A. Kulesza, G.F. Sterman and W. Vogelsang,
\newblock Phys.Rev. D66 (2002) 014011.

\bibitem{Banfi:2010cf}
A. Banfi, S. Redford, M. Vesterinen, P. Waller and T.R. Wyatt,
\newblock Eur.Phys.J. C71 (2011) 1600.

\bibitem{Balazs:1997xd}
C. Balazs and C.-P. Yuan,
\newblock Phys.Rev. D56 (1997) 5558.

\bibitem{RESBOSweb}
\textsc{ResBos} is available at http://hep.pa.msu.edu/resum/

\bibitem{Banfi:2012du}
A. Banfi, M. Dasgupta, S. Marzani and L. Tomlinson.
\newblock Phys.Lett. 715 (2012) 152-156.

\bibitem{Banfi:2011dx}
A. Banfi, M. Dasgupta and S. Marzani,
\newblock Phys.Lett. B701 (2011) 75.

\bibitem{Banfi:2011dm}
A. Banfi, M. Dasgupta, S. Marzani and L. Tomlinson.
\newblock JHEP 1201 (2012) 044.

\bibitem{Banfi:2009dy}
A. Banfi, M. Dasgupta and R.M. Duran~Delgado,
\newblock JHEP 0912 (2009) 022.

\bibitem{Baur:2001ze}
U. Baur, O. Brein, W. Hollik, C. Schappacher and D. Wackeroth,
\newblock Phys.Rev. D65 (2002) 033007.

\bibitem{Zykunov:2005tc}
V. Zykunov,
\newblock Phys.Rev. D75 (2007) 073019.

\bibitem{CarloniCalame:2007cd}
C.M. Carloni Calame, G. Montagna, O. Nicrosini and A. Vicini,
\newblock JHEP 0710 (2007) 109.

\bibitem{Arbuzov:2007db}
A. Arbuzov, D. Bardin, S. Bondarenko, P. Christova, L. Kalinovskaya, G. Nanava and R. Sadykov
\newblock Eur.Phys.J. C54 (2008) 451.

\bibitem{Barberio:1993qi}
E. Barberio and Z. Was,
\newblock Comput.Phys.Commun. 79 (1994) 291.

\bibitem{Arnold:1989ub}
P.B. Arnold, R.K. Ellis and M. Reno,
\newblock Phys.Rev. D40 (1989) 912.

\bibitem{Arnold:1988dp}
P.B. Arnold and M.H. Reno,
\newblock Nucl.Phys. B319 (1989) 37.

\bibitem{Melnikov:2006kv}
K. Melnikov and F. Petriello,
\newblock Phys.Rev. D74 (2006) 114017.

\bibitem{Hamberg:1990np}
R. Hamberg, W. van Neerven and T. Matsuura,
\newblock Nucl.Phys. B359 (1991) 343.

\bibitem{Cafarella:2007tj}
A. Cafarella, C. Corian\`o and M. Guzzi,
\newblock JHEP 0708 (2007) 030.

\bibitem{Cafarella:2008du}
A. Cafarella, C. Corian\`o and M. Guzzi,
\newblock Comput.Phys.Commun. 179 (2008) 665.

\end{thebibliography}

\end{document}